\DocumentMetadata{}
\documentclass[manuscript,screen,review=false]{acmart}

\renewcommand\footnotetextcopyrightpermission[1]{} 
\theoremstyle{thmstyleone}%
\usepackage{natbib}
\newtheorem{theorem}{Theorem}

\theoremstyle{thmstyletwo}%
\theoremstyle{thmstylethree}%
\newtheorem{definition}{Definition}%
\usepackage{algorithm}
\usepackage{algorithmic}
\usepackage[utf8]{inputenc}
\usepackage[english]{babel}
\usepackage{cancel}
\usepackage{amsmath}
\usepackage{url}
\usepackage{multirow}
\usepackage{array,multirow}
\usepackage{algorithm}
\usepackage{graphicx}
\usepackage{verbatim}
\graphicspath{ {./figure/} }
\addto\captionsenglish{}
\usepackage{babel}
\usepackage{xfrac}
\usepackage{dblfloatfix}

\usepackage{etoolbox}
\makeatletter
\patchcmd{\@makecaption}
  {\scshape}
  {}
  {}
  {}
\makeatother

\usepackage{xcolor}
\usepackage{tabularx}
\usepackage{amsthm}
\usepackage{array, boldline, makecell, booktabs}

\newtheorem{lemma}{Lemma}
\renewenvironment{proof}{{\bfseries Proof.}}{\qed}
\usepackage{multirow}
\usepackage{enumitem}
\usepackage{diagbox}
\usepackage{afterpage}

\usepackage{color,soul}
\usepackage{array,multirow}
\usepackage{caption}

\usepackage{microtype}
\usepackage{booktabs}
\usepackage[utf8]{inputenc}

\AtBeginDocument{%
  \providecommand\BibTeX{{%
    \normalfont B\kern-0.5em{\scshape i\kern-0.25em b}\kern-0.8em\TeX}}}

\setcopyright{acmcopyright}
\copyrightyear{2024}
\acmYear{2024}
\acmDOI{XXXXXXX.XXXXXXX}

\acmConference[ACM Transactions on the Web]
{Make sure to enter the correct conference title from your rights confirmation emai}{June 03--05,2024}{Woodstock, NY}
\acmPrice{15.00}
\acmISBN{978-1-4503-XXXX-X/18/06}

\begin{document}

\title{Tracking Top-\textit{k} Structural Hole Spanners in Dynamic Networks\\ \vspace{0.2in}
\centering
\textnormal{Diksha Goel$^{1,2}$, Hong Shen$^{3}$, Hui Tian$^{4}$, Mingyu Guo$^{1}$\\ \vspace{0.1in} \textnormal{$^{1}$University of Adelaide, Australia} \\
\textnormal{$^{2}$CSIRO’s Data61, Australia} \\
\textnormal{$^{3}$Central Queensland University, Australia} \\ \textnormal{$^{4}$Griffith University, Australia}}\\
\small \textnormal{diksha.goel@data61.csiro.au, h.shen@cqu.edu.au, hui.tian@griffith.edu.au, mingyu.guo@adelaide.edu.au}}

\renewcommand{\shortauthors}{Goel et al.}


\begin{abstract}

Structural Hole (SH) theory states that the node which acts as a connecting link among otherwise disconnected communities gets positional advantages in the network. These nodes are called Structural Hole Spanners (SHS). Numerous solutions are proposed to discover SHSs; however, most of the solutions are only applicable to static networks. Since real-world networks are dynamic networks; consequently, in this study, we aim to discover SHSs in dynamic networks. We first propose an efficient Tracking-SHS algorithm for updating SHSs in dynamic networks. Our algorithm reuses the information obtained during the initial runs of the static algorithm and avoids the recomputations for the nodes unaffected by the updates. Besides, we also design a Graph Neural Network-based model, GNN-SHS, to discover SHSs in dynamic networks, aiming to reduce the computational cost while achieving high accuracy. We provide a theoretical analysis of the Tracking-SHS algorithm and prove that the Tracking-SHS algorithm achieves 1.6 times speedup compared with the static algorithm. We perform extensive experiments, and our results demonstrate that the Tracking-SHS algorithm attains a minimum of 3.24 times speedup over the static algorithm. Also, the proposed second model GNN-SHS, is, on average 671.6 times faster than the Tracking-SHS algorithm.

\end{abstract}

\maketitle
\pagestyle{plain}
\section{Introduction}\label{sec1}

\noindent The emergence of large-scale networks has inspired researchers to design new techniques to analyze and study the properties of these large-scale networks \cite{guo2021itcn, binesh2021distance}. The structure of these network inherently possesses a community structure where the nodes within the community share close interests, characteristics, behaviour, and opinions \cite{zannettou2018origins}. The absence of connection between different communities in the network is known as \textbf{\textit{Structural Holes (SH)}} \cite{burt2009structural}. A community needs to have connectivity with other communities in order to have access to the novel information \cite{rinia2001citation}.

\begin{figure}[!ht]
  \centering
    \includegraphics[width=0.28\paperwidth]{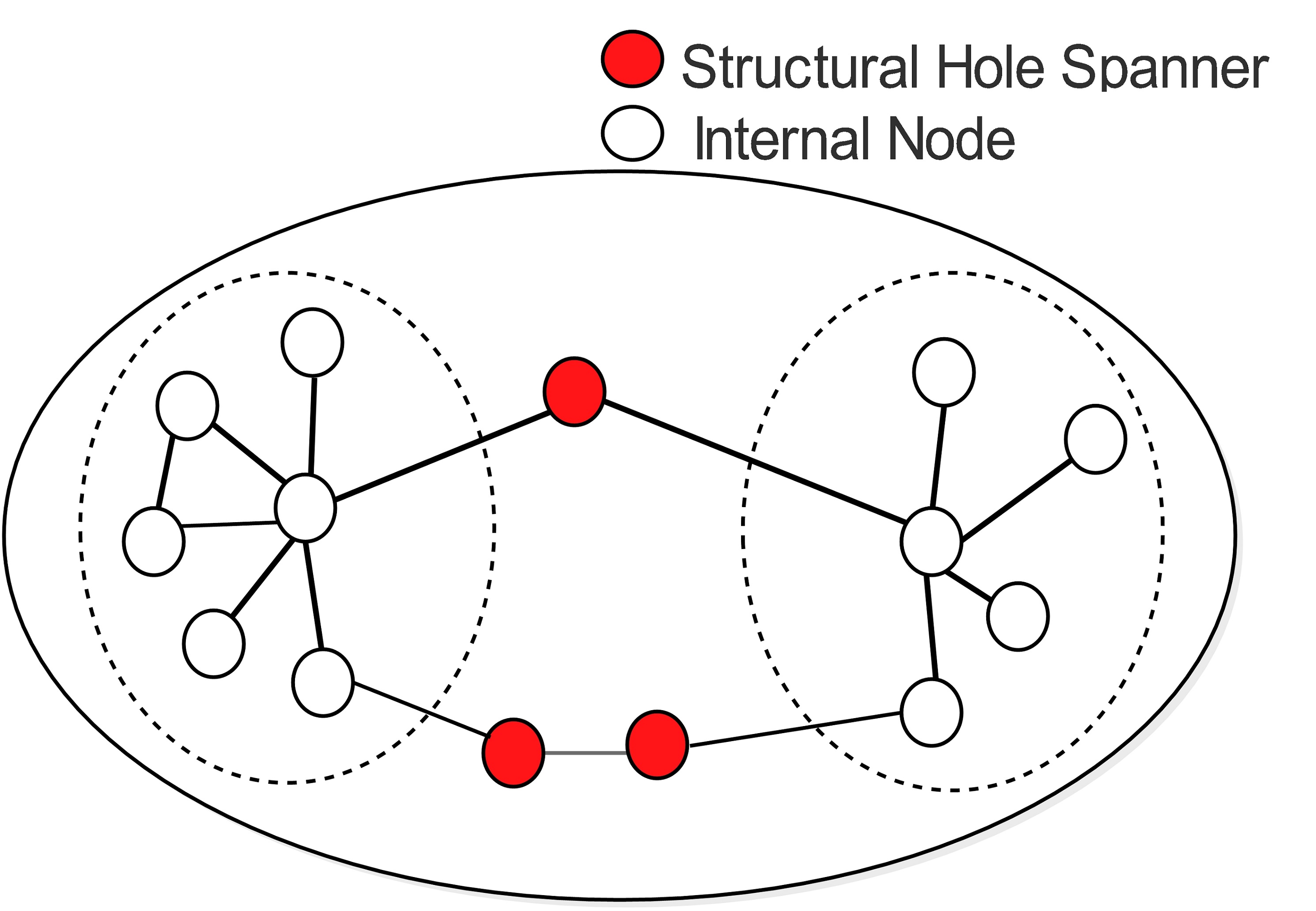}
    \caption{Structural Hole Spanners in the network.}
    \label{fig:SHS}
\end{figure}

The structural hole theory states that the users who fill the “\textit{holes}” between various users or groups of users that are otherwise disconnected get positional advantages in the network. These users are known as \textbf{\textit{Structural Hole Spanners}} \cite{lou2013mining}. Figure \ref{fig:SHS} illustrates SHSs in the network. SHSs have many applications, including information diffusion,
opinion control, identifying central hubs, preventing the spread of rumours, community detection and identifying critical
nodes in tactical environments \cite{lin2021efficient, amelkin2019fighting, zareie2019influential, yu2021modeling, zhao2021community, goel2018overview, abdulsatar2024towards}

\begin{figure*}[!t]
  \centering
    \includegraphics[width=0.58\paperwidth]{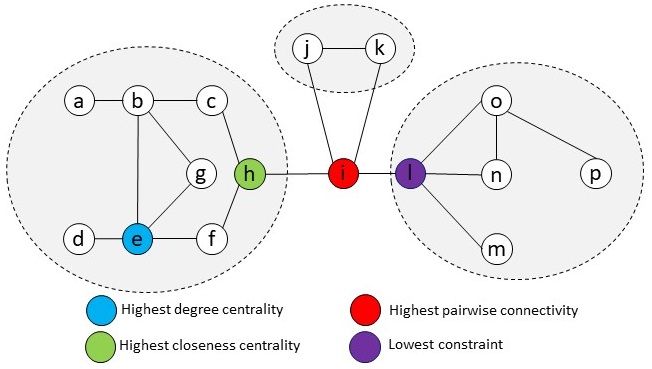}
    \caption{Comparison of various centrality measures.}
    \label{fig:comp}
  \end{figure*}

A large number of centrality measures, such as pairwise connectivity \cite{borgatti2006identifying}, closeness centrality \cite{bavelas1950communication}, degree centrality, constraint \cite{burt1992structural} etc., exist in literature to discover critical nodes in the network.
However, SHS acts as a bridge between the nodes of different communities \cite{rezvani2015identifying} and controls information diffusion in the network \cite{lou2013mining}. Therefore, we have an important observation that the removal of SHS will disconnect a maximum number of node pairs in the network and block information propagation among them. Figure \ref{fig:comp} illustrates that node $i$ plays a vital role, and removing node $i$ will disconnect maximum node pairs in the network, in turn blocking information propagation among the maximum number of nodes in the network, whereas the impact of the removal of other nodes is less significant. Therefore, 
We define SHS as a node whose removal minimizes the \textit{Pairwise Connectivity (PC)} of the residual network, i.e., the node with the maximum pairwise connectivity score. This SHS definition aims to capture the nodes that are located between otherwise disconnected groups of nodes.

Various solutions \cite{lou2013mining, rezvani2015identifying, gong2019identifying, goyal2007structural, he2016joint, xu2017efficient, xu2019identifying, tang2012inferring, ding2016method} are developed to discover SHSs in static networks. Nevertheless, the real-world network changes over time. For example, it is essential to handle outdated web links in web graphs or obsolete user profiles in a social network; these are examples of decremental updates in the networks. In contrast, on Facebook and Twitter, links appear and disappear whenever a user friend/unfriend others on Facebook or follow/unfollow others on Twitter, which is an example where decremental as well as incremental updates happen in the networks. As a result, discovered SHSs change. \textit{The limitation here is that in the literature, there is no solution available to discover SHS nodes in dynamic networks.} The classical SHS identification algorithms are considerably time-consuming and may not work efficiently for dynamic networks. In addition, the network may have already been changed by the time classical algorithms recompute SHSs. Hence, it is crucial to design a fast mechanism that can efficiently discover SHSs as the network evolves \cite{goel2023enhancing}.  

\textit{In this paper\footnote{This paper is an extension of our prior work \cite{goel2021maintenance} and \cite{goel2022discovering}, accepted at LCN 2021 and WI-IAT 2022 respectively.}, we aim to propose efficient algorithms for discovering structural hole spanner nodes in dynamic networks.} 
We formulate the problem of discovering SHS nodes in dynamic networks as \textbf{\textit{Structural hole Spanner Tracking (SST) problem}}. While the traditional SHS problem focuses on discovering a set of SHSs that minimize the PC of the network, the SST problem intends to update the already discovered set of SHSs as the network evolves. 

In order to track SHS nodes in dynamic networks, we first propose an efficient \textbf{\textit{Tracking-SHS algorithm}} that maintains Top-$k$ SHSs for decremental edge updates in the network by discovering a set of affected nodes. \textit{Tracking-SHS aims to maintain and update the SHS nodes faster than recomputing them from the ground.} We derive some properties to determine the affected nodes due to updates in the network. In addition, we reuse the information from the initial runs of the static algorithm in order to avoid the recomputations for the unaffected nodes. Tracking-SHS executes greedy interchange by replacing an old SHS node with a high PC score node from the network. Besides, the use of priority queues saves the repetitive computation of PC scores. Our theoretical and empirical results demonstrate that the proposed Tracking-SHS algorithm achieves higher speedup than the static algorithm.

In addition, inspired by the recent advancements of Graph Neural Network on various graph problems, we propose another model \textbf{\textit{GNN-SHS}} (\underline{G}raph \underline{N}eural \underline{N}etworks for discovering \underline{S}tructural \underline{H}ole \underline{S}panners in dynamic networks), a graph neural network-based framework that discovers SHSs in the dynamic network. In our proposed model, we consider both incremental as well as decremental edge updates of the network. We regard the dynamic network as a sequence of snapshots and aim to discover SHSs in these snapshots. Our proposed  GNN-SHS model uses the network structure and features of nodes to learn the embedding vectors of nodes. Our model aggregates embedding vectors from the neighbours of the nodes, and the final embeddings are used to discover SHS nodes in the network. GNN-SHS aims to discover SHSs in dynamic networks by reducing the computational time while achieving high accuracy.

{We perform a detailed theoretical analysis of the proposed Tracking-SHS algorithm, and our results show that for a specific type of graph, the Tracking-SHS algorithm achieves a 1.6x speedup compared to the static algorithm. Tracking-SHS is designed as a deterministic, algorithmic approach specifically optimized for decremental-only updates in dynamic networks. Through worst-case performance analysis, we demonstrate that Tracking-SHS provides consistent, predictable efficiency, making it highly suitable for applications dominated by decremental updates, such as managing outdated links or handling social network disconnections. This theoretical foundation offers a reliable benchmark for decremental-case scenarios, assuring performance consistency in strictly defined dynamic environments. In contrast, the GNN-SHS model is a machine learning-based approach leveraging Graph Neural Networks to handle both decremental and incremental updates. This learning-based model provides adaptability across a variety of network update types, allowing it to capture complex, non-linear patterns in dynamic networks. GNN-SHS is trained to generalize across both types of updates, making it valuable for real-world applications where network connections evolve unpredictably and both additions and deletions are frequent. By adapting to diverse changes through learned embeddings, GNN-SHS provides a scalable solution. We validated the performance of both the Tracking-SHS algorithm and the GNN-SHS model by conducting extensive experiments on real-world and synthetic datasets. Our experimental results show that the Tracking-SHS algorithm is at least 3.24 times faster than the static algorithm, verifying its efficiency in decremental-update environments. In addition, the GNN-SHS model achieves exceptional speedups, ranging from 31.8x to as high as 2996.9x faster than the baseline, underscoring its considerable efficiency advantage in handling both types of updates in dynamic networks.}

\noindent Our contributions can be summarized as follows:
\begin{itemize}[leftmargin=0.4cm]
\item  \textbf{Tracking-SHS algorithm.} We propose an efficient algorithm, Tracking-SHS, that maintains SHS nodes for decremental edge updates in the network. We derive some properties to discover the set of nodes affected due to updates in the network and avoid recomputation for the unaffected nodes, so as to enhance the efficiency of the proposed algorithm. In addition, we extend our proposed algorithm from a single edge update to a batch of updates.
\item \textbf{GNN-SHS model.} We propose an efficient graph neural network-based model, GNN-SHS, for discovering SHSs in dynamic networks. Our model considers both the incremental and decremental edge updates in the network. The GNN-SHS model preserves the network structure and node feature and uses the final node embedding to discover SHSs.
\item \textbf{Theoretical analysis.} We theoretically show that the depth of the proposed graph neural network-based model GNN-SHS should be at least $\Omega({n}^2/\log^2 n)$ to solve the SHSs problem. In addition, we analyze the performance of the proposed Tracking-SHS algorithm theoretically, and our theoretical results show that for specific types of graphs, such as Preferential Attachment graphs, the proposed algorithm achieves 1.6 times of speedup compared with the static algorithm.
\item \textbf{Experimental analysis.} We validate the performance of the proposed Tracking-SHS algorithm and GNN-SHS model by conducting extensive experiments on various real-world and synthetic datasets. The results demonstrate that the Tracking-SHS algorithm achieves a minimum of 3.24 times speedup on real-world datasets compared with the static algorithm. Besides, the GNN-SHS model is at least 31.8 times faster and, on average, 671.6 times faster than the Tracking-SHS algorithm.
\end{itemize}
\noindent \textbf{Organization.} Section \ref{sec2} discusses the related work. Section \ref{sec3} presents the preliminaries and problem formulation. Section \ref{sec4} discusses the proposed Tracking-SHS algorithm for tracking SHSs in dynamic networks. Section \ref{sec5} discusses the proposed model GNN-SHS for discovering SHSs in dynamic networks. Section \ref{sec6} presents the theoretical performance analysis of the Tracking-SHS algorithm, and Section \ref{sec7} discusses the extensive experimental results. Finally, Section \ref{sec8} concludes the paper.

\section{Related Work}\label{sec2}
\noindent Identification of SHSs has numerous applications, and various studies \cite{lou2013mining, rezvani2015identifying, gong2019identifying, goyal2007structural, he2016joint, xu2017efficient, xu2019identifying, tang2012inferring, ding2016method} have been conducted to discover SHS nodes. Existing work on SHSs identification problems can be categorized into 1) discovering SHSs in static networks; and 2) discovering SHSs in dynamic networks. This section discusses the state-of-the-art for discovering SHSs. 

\subsection{Discovering SHSs in Static Networks} 
\noindent This section discusses the various solutions for discovering SHSs in static networks. We further categorized the SHSs solutions into information propagation-based and network centrality-based solutions. Table \ref{related work} shows the summary of SHS identification solutions for static networks.\\

\noindent \textbf{Information propagation based solutions\\} 
\noindent Lou et al. \cite{lou2013mining} proposed a technique based on minimal cut to discover SHSs in the network. However, the technique requires community information in the network, and the quality of identified SHS depends on how accurately communities are identified. He et al. \cite{he2016joint} developed a solution to identify SHSs and communities in the network. The authors explored the types of interactions, particularly for bridging nodes, in order to distinguish SHSs from the other nodes. The model assumed that every node is associated with one community only; however, in the real world, a node may belong to multiple communities \cite{yang2015defining}. Xu et al. \cite{xu2019identifying} proposed the maxBlock algorithm to discover SHSs and argued that SHSs are the nodes that block maximum information propagation when removed from the network. Gong et al. \cite{gong2020structural} proposed a mechanism for controlling the opinion of the public using SHS nodes. The goal of this approach is to reduce opinion polarity in the network. \\

\noindent \textbf{Network centrality based solutions\\} 
\noindent Tang et al. \cite{tang2012inferring} developed a 2-step solution based on shortest paths of length two to discover SHSs. Rezvani et al. \cite{rezvani2015identifying} developed several solutions based on closeness centrality to discover SHSs. Xu et al. \cite{xu2017efficient} proposed fast and scalable solutions for discovering SHS nodes. Ding et al. \cite{ding2016method} devised a V-Constraint for discovering critical nodes that fill SH in the network. The model considered various node features, such as its neighbours' degree and topological features, to identify the SHS nodes. However, the local features of a node may not capture the global importance of the SHS node. Goyal et al. \cite{goyal2007structural} designed the SHS problem as a set of nodes that pass through many shortest paths between different pair of nodes. However, it may take a significant amount of time to determine all pair shortest paths for a large network. Zhang et al. \cite{zhang2020finding} used a community forest model to identify SHS nodes. The authors argued that local features-based metrics cannot identify SHSs in the network. Luo et al. \cite{luo2020detecting} designed a deep learning model, ComSHAE, that identifies SHSs and communities in the network. Goel et al. \cite{goel2024effective} addressed the challenges in SHS discovery for large-scale and diverse networks by introducing two GNN-based models: GraphSHS and Meta-GraphSHS. GraphSHS optimizes SHS discovery by leveraging network structure and node features to reduce computational costs. Meta-GraphSHS, using meta-learning, learns transferable knowledge from diverse graphs, allowing it to adapt effectively to new networks with high accuracy.

\begin{table*}[!t] 
\caption{Summary of SHSs identification solutions for static networks.}
\label{related work}
\renewcommand{\arraystretch}{1.2}
\centering 
\begin{tabular}{l p{2.2cm} p{3.5cm} p{4cm}} \toprule
\textbf{Category of solution} & \textbf{Author} & \textbf{Algorithm} & \textbf{Pros \& Cons} \\ \hlineB{1.5}

\multirow{4}{*}{Information propagation} &   Lou et al. \cite{lou2013mining} & HIS, MaxD & Require prior community information \\ \cmidrule{2-4}
&  He et al. \cite{he2016joint} & HAM & Jointly discover SHSs and communities \\ \cmidrule{2-4}
&  Xu et al. \cite{xu2019identifying} & maxBlock, maxBlockFast & Less computational cost \\\cmidrule{2-4}
&  Gong et al. \cite{gong2019identifying} &Machine learning model   & Achieves high accuracy \\\hlineB{1.5}

\multirow{10}{*}{Network centrality} &   Tang et al. \cite{tang2012inferring} & 2-step algorithm &  Fails if node is densely connected to many communities\\ \cmidrule{2-4}

& Rezvani et al. \cite{rezvani2015identifying} & ICC, BICC, AP\_BICC & Only used topological network structure \\ \cmidrule{2-4} 

 &   Xu et al. \cite{xu2017efficient} &Greedy, AP\_Greedy & Does not require community information\\ \cmidrule{2-4}
 
 &   Ding et al. \cite{ding2016method} & V-Constraint & Ego network may not capture the global importance of the node \\ \cmidrule{2-4}
 
  &   Goyal et al. \cite{goyal2007structural} & Network formation model & High computational cost\\ \cmidrule{2-4}

  &   Zhang et al. \cite{zhang2020finding} & FSBCDM & Local features cannot discover SHSs \\\cmidrule{2-4}
  
  &  Luo et al. \cite{luo2020detecting} & ComSHAE & Jointly discover SHSs and communities\\ \hlineB{1.5}
\end{tabular} 
\end{table*}

\subsection{Discovering SHSs in Dynamic Networks} 
\noindent Numerous solutions have been proposed for discovering SHSs for the steady-state behaviour of the network. However, real-world networks are not static; they evolve continuously.\textit{ Currently, there is no solution that discovers SHS nodes in dynamic networks.} Unlike previous studies, this paper aims to discover SHS nodes in dynamic networks.

\section{Preliminaries and Problem Formulation}\label{sec3}
\noindent This section first discusses the preliminaries and background of the problem. Later, we formally state the structural hole spanner problem for the static network and the structural hole spanner problem for the dynamic network.

\subsection{Preliminaries and Background}
\noindent \textbf{Network model.}
\noindent A graph can be defined as $G = (V, E)$, where $V$ is the set of nodes (vertices), and $E \subseteq V \times V$ is the set of edges. Let $n = |V |$ and $m = |E|$. We have considered unweighted and undirected graphs without self-loops or multiple edges. A \textit{path} $p_{ij}$ from node $i$ to $j$ in an undirected graph $G$ is a sequence of nodes ${\{v_{i},v_{i+1},.....,v_{j}\}}$  such that each pair  $(v_{i},v_{i+1})$ is an edge in $E$. A pair of nodes $i$, $j \in V$ is \textit{connected} if there is a path between $i$ and $j$. Graph $G$ is connected if all pair of nodes in $G$ are connected. Otherwise, $G$ is \textit{disconnected}. A \textit{connected component} or \textit{component} $C$ in an undirected graph $G$ is a maximal set of nodes in which a path connects each pair of node.

\begin{table}[!t] 
\caption{The summary of symbols.}
\label{notations}
\renewcommand{\arraystretch}{1.25}
\centering 
\begin{tabular}{l l} \hlineB{1.5} 
\textbf{Symbol} & \textbf{Description} \\ \hlineB{1.5}
$G$, $G'$ & Original and Updated graph \\ 
$G_t$ & Snapshot of graph at time $t$ \\
$n,m, k$ & Number of nodes, edges and SHSs\\  
$m(i)$ & Number of edges in component containing node $i$ \\ 
$P(G)$ & Pairwise connectivity of graph \\ 
$P(G\backslash\{i\})$ & Pairwise connectivity of graph without node $i$ \\
$p_{ij}$ & Path from node $i$ to $j$\\ 
$u(i,j)$ & Pairwise connectivity between node $i$ and $j$ \\ 
$c(i)$ & Pairwise connectivity score of node $i$ \\ 
$c'(i)$ & Pairwise connectivity score of $i$ in updated graph\\
$C$ & Connected component\\ 
$C(i)$ & Connected component containing node $i$ \\
$ \mid C(i) \mid $ & Number of nodes in component containing node $i$ \\
$N(i)$ & Neighbors of node $i$ \\
$ \mid N(i) \mid $ & Number of neighbors of node $i$ \\
$l$ & Index of aggregation layer \\ 
$z(i)$ & Final embeddings of node $i$ \\
$h^{l}(i)$ & Embeddings of node $i$ at the $l^{th}$ layer \\ 
$y(i)$ & Label of node $i$ \\
$\vec{x}(i)$ & Feature vector of node $i$\\ \hlineB{1.5} 
\end{tabular} 
\end{table}

A dynamic graph $G$ can be modelled as a finite sequence of graphs $(G_t, G_{t+1}, ..., G_T)$. Each $G_t$ graph represents the network's state at a discrete-time interval $t$. We refer to each of the graph in the sequence as a snapshot graph. Each snapshot consists of the same set of nodes, whereas edges may appear or disappear over time. Hence, each graph snapshot can be described as an undirected graph $G_t = (V, E_t)$, containing all nodes and only alive edges at the time interval under consideration. Due to the dynamic nature of the graph, the edges in the graph may appear or disappear, due to which the label of the nodes (SHS or normal node) may change. Therefore, we need to design techniques that can discover SHSs in each new snapshot graph quickly.

Let $\vec{x}(i)$ denotes the feature vector of node $i$, $N(i)$ denotes the neighbours of node $i$, $h^{l}(i)$ represents the embedding of node $i$ at the $l^{th}$ layer of the model. Let $l$ represents the index of the aggregation layer, where $l = (1,2,...,L)$. Table \ref{notations} presents the symbols used in this paper.

\begin{definition}
\noindent \textbf{Pairwise connectivity.} \normalfont The pairwise connectivity $u(i,j)$ for any node pair $(i,j)\in V \times V $ is quantified as: 
{\begin{equation}
u(i,j)=\begin{cases} 1 & \text{if\,$i$\,and\,$j$\,are\,connected} \\ 0 & \text{otherwise}
\end{cases} 
\end{equation}}
\end{definition}

\begin{definition}
\noindent  \textbf{Total pairwise connectivity.}  \normalfont The total pairwise connectivity $P(G)$, i.e., pairwise connectivity across all pair of nodes in the graph, is given by:
{\begin{equation}
\text{\ensuremath{{\displaystyle P(G)=\sum_{i,j\in V\times V,i\neq j}{\textstyle u(i,j)}}}}
\end{equation}}
\end{definition}
\begin{definition}

\noindent  \textbf{Pairwise connectivity score.}  \normalfont The pairwise connectivity score $c(i)$ of node $i$ is the contribution of node $i$ to the total pairwise connectivity score of the graph. Pairwise connectivity score $c(i)$ of node $i$ is computed as follows:
{\begin{equation}
\label{pc_score}
c(i)=P(G)-P(G\backslash\{i\})
\end{equation}}
\end{definition}

\noindent \textbf{Graph Neural Networks (GNNs).} GNNs \cite{wu2020comprehensive} are designed by extending the deep learning methods for the graph data and are broadly utilized in various fields, e.g., computer vision, graph mining problems, etc. GNNs usually consist of graph convolution layers that extract local structural features of the nodes  \cite{velivckovic2017graph}. GNNs learn the representation of nodes by aggregating features from the ego network of node. Every node in network is described by its own features, and features of its neighbors \cite{kipf2016semi}.\\

\begin{figure}[t!]
 \centering
 \includegraphics[width=0.53\columnwidth]{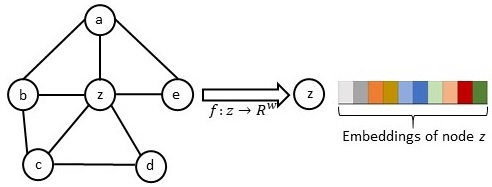}
 \caption{Embedding of node $i$.}
 \label{fig:embed}
\end{figure}

\noindent \textbf{Network Embedding.} Network embedding \cite{cui2018survey} is a procedure with which network nodes can be described in a low-dimensional vector. Embedding intends to encode the nodes so that the resemblance in the embedding approximates the resemblance in the network \cite{aguilar2021novel}. Embedding can be used for various graph mining problems, including node classification, regression problems, graph classification, etc. Figure \ref{fig:embed} depicts an example of node embedding in low-dimensional space.

\subsection{{Problem Statement}}
\subsubsection{Structural Hole Spanner Problem for Static Network}

\noindent This section discusses the greedy algorithm for discovering SHSs in the static network. Selecting a set of SHS nodes in one observation may not be a correct approach due to the influence of cut nodes. Therefore, the algorithm works iteratively by identifying one SHS node in each iteration.

\begin{definition}
\noindent \textbf{Structural Hole Spanner Problem.}  \normalfont Given a graph $G = (V,E)$, and a positive integer $k$, the \textit{SHS problem} is to identify a set of SHSs Top-$k$ in $G(V, E)$, where Top-$k$ $\subset V$ and $ \mid $Top-$k \mid = k$, such that the removal of nodes in Top-$k$ from $G$ minimizes the total pairwise connectivity in the residual subgraph $ G(V\backslash $Top-$k)$.
\end{definition}
\begin{equation}
\text{Top-}k = {min}\,\,\{{P(G\backslash \text{Top-}k)}\}
\end{equation}
where Top-$k\,\subset\,V$ and $ \mid $Top-$k \mid =k$.\\

Algorithm \ref{alg 1} describes a greedy heuristic for identifying the SHSs in the static networks. The algorithm repeatedly selects a node $v'$ with a maximum PC score, i.e., the node which when removed from the network minimizes the total PC of the residual network. In step 3, for computing the PC score of each node $v \in V$, the algorithm initiates a depth-first search (DFS) and selects a node with the maximum PC score. The selected node is then eliminated from the network and added to the SHS set Top-$k$. The process repeats until $k$ nodes are discovered. The run time of Algorithm \ref{alg 1} is $O(kn(m+n))$. DFS takes $O(m+n)$ time and the PC score is calculated for $n$ nodes. The process repeats for $k$ iterations, and hence, the complexity follows.

\begin{algorithm}[h!]
 \caption{Structural hole spanner identification.}
 \label{alg 1}
 \begin{algorithmic}[1]
 \renewcommand{\algorithmicrequire}{\textbf{Input:}}
 \renewcommand{\algorithmicensure}{\textbf{Output:}}
 \REQUIRE Graph $G(V, E)$, $k$
 \ENSURE  SHS set Top-$k$
  \STATE Initialize Top-$k=\phi$
  \WHILE{$|$Top-$k|<k$}
  \STATE $v'=argmax_{v\,\text{\ensuremath{\in}}\,V}\,c(v)$
  \STATE $G=G\backslash\{v'\}$
  \STATE Top-$k=$Top-$k\bigcup\{v'\}$
  \ENDWHILE
 \RETURN Top-$k$
 \end{algorithmic}
 \end{algorithm}

\begin{theorem} [\textbf{Dinh et al. \cite{dinh2011new}}] \label{thm1}
\textit{\textbf{Discovering SHS problem is NP-hard.}}
\end{theorem}

\noindent \begin{proof}
We present an alternative proof, where we reduce the SHS model to vertex cover instead of $\beta$-Vertex Disruptor used in Dinh et al. \cite{dinh2011new}. The reason for this alternative proof is that it will be used as a foundation for Theorem 2. 

We show that the Vertex Cover (VC) problem is reducible to the SHS discovery problem. The definition of \textit{Structural Hole Spanners} states that SHSs are the set of $k$ nodes,  which, when deleted from the graph, minimizes the total pairwise connectivity of the subgraph. Let $G=(V, E)$ be an instance of a VC problem in an undirected graph $G$ with $V$ vertices and $E$ edges. \textit{VC problem} aims to discover a set of vertices of size $k$ such that the set includes at least one endpoint of every edge of the graph. If we delete the nodes in vertex cover from the graph, there will be no edge in the graph, and the pairwise connectivity of the residual graph will become $0$, i.e., $P(G)$ is minimized. In this way, we can say that graph $G$ has a VC of size $k$ if and only if graph $G$ have structural hole spanners of size $k$. Therefore, discovering the exact $k$ SHSs problem is NP-hard as a similar instance of a vertex cover problem is a known NP-hard problem.
\end{proof}

\subsubsection{Structural Hole Spanner Problem for Dynamic Network}

\begin{figure}[t!]
 \centering
 \includegraphics[width=0.55\columnwidth]{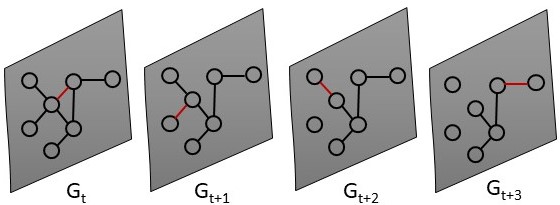}
 \caption{Illustration of snapshots of graph.}
 \label{fig:evolving}
\end{figure}

\noindent We represent the dynamic network as a sequence of snapshots of graph $(G_t, G_{t+1}, ..., G_T)$ and each snapshot graph describes all the edges that occur between a specified discrete-time interval (e.g., sec, minute, hour). Figure \ref{fig:evolving} illustrates four snapshots of graph taken at time $t, t+1, t+2$, $t+3$. Due to the dynamic nature of the graph, SHSs in the graph also change, and it is crucial to discover SHSs in each new snapshot graph quickly. Traditional SHSs techniques are time-consuming and may not be suitable for dynamic graphs. Therefore, we need fast algorithms to discover SHSs in dynamic networks. We formally define the structural hole spanner problems for dynamic networks as follow:\\

\noindent \textbf{{Problem 1:}} \textit{\textbf{Tracking Top-$k$ SHS nodes in Dynamic Networks (Structural Hole Spanner Tracking Problem).}}

\noindent \textbf{{Given:}}   \normalfont Given a graph $G = (V, E)$, SHS set Top-$k$, and edge update $\Delta E$.

\noindent \textbf{\textit{Goal:}} Design a Tracking-SHS algorithm to identify SHS set Top$'$-$k$ with cardinality $k$ in $G'=(V,E + \Delta E)$ by updating Top-$k$ such that the removal of nodes in Top$'$-$k$ from $G'$ minimizes the total pairwise connectivity in the residual subgraph $G'(V\backslash$Top$'$-$k$).\\

\noindent \textbf{{Problem 2:}} \textit{\textbf{Discovering Top-$k$ SHS nodes in Dynamic Networks.}}

\noindent \textbf{{Given:}} Given snapshots of graph $G_t = (V,E_t), G_{t+1} = (V,E_{t+1})$,...$G_{T} = (V,E_{T})$ and integer $k > 0$.

\noindent \textbf{\textit{Goal:}} Train a model GNN-SHS to discover SHS set Top-$k$  with cardinality $k$ in dynamic network (snapshots of graph) such that the removal of nodes in Top-$k$ from the snapshot of graph minimizes the total pairwise connectivity in the residual subgraph. We aim to utilize the pre-trained model to discover SHSs in each new snapshot of the graph quickly. \\

Theorem 1 showed that discovering the exact $k$ SHSs in the network is an NP-hard problem. Therefore, we adopted a greedy heuristic, as discussed in Algorithm 1 for finding the Top-$k$ SHS nodes. It should be noted that the greedy algorithm, despite being a heuristic, is still computationally expensive with a complexity of  $O(kn(m+n))$. Nevertheless, real-world networks are dynamic, and they change rapidly. Since these networks change quickly,  the Top-$k$ SHSs in the network also change continuously. A run time of $O(kn(m+n))$ of the greedy algorithm is too high and not suitable for practical purposes where speed is the key. For instance, it is highly possible that the network might have already changed by the time greedy algorithm computes the Top-$k$ SHSs. Therefore, we need an efficient solution that can quickly discover Top-$k$ SHSs in the changing networks.

A natural approach for the SST problem is to run Algorithm \ref{alg 1} after each update, providing us with a new set of SHSs. Nevertheless, computing SHS set from scratch after every update is a time-consuming process which motivates us to design \textit{Tracking-SHS algorithm} that can handle the dynamic nature of the networks. Tracking-SHS algorithm focuses on updating the Top-$k$ SHSs without explicitly running Algorithm \ref{alg 1} after every update.

We also propose a GNN-based model to discover SHSs in dynamic networks by transforming the SHS discovery problems into a learning problem. While our proposed Tracking-SHS algorithm only considers decremental edge updates in the network, our second proposed model, GNN-SHS considers both incremental as well as decremental edge updates of the network. The main idea of our second model \textit{GNN-SHS}, is to rely on the greedy heuristic (Algorithm \ref{alg 1}) and treat its results as true labels to train a graph neural network for identifying the Top-$k$ SHSs. The end result is a significantly faster heuristic for determining the Top-$k$ SHSs. Our heuristic is faster because we only have to train our graph neural network model once, and thereafter, whenever the graph changes, the trained model can be utilized to discover SHSs. Table \ref{inc_dec} presents the comparison of both the proposed models for discovering SHS nodes in dynamic networks.

\begin{table}[!t] 
\caption{Comparison of both the proposed models for discovering SHS nodes in dynamic networks.}
\label{inc_dec}
\renewcommand{\arraystretch}{1.15}
\centering 
\begin{tabular}{p{2.5cm}p{1.7cm}p{1.7cm}p{1.2cm}} \hlineB{1.5} 
\textbf{Proposed algorithm/model} & \textbf{Decremental update} & \textbf{Incremental update} & \textbf{Batch update}\\ \hlineB{1.5}
Tracking-SHS & \checkmark &   $\times$   & \checkmark\\
GNN-SHS & \checkmark & \checkmark & \checkmark\\ \hlineB{1.5} 
\end{tabular} 
\end{table}

\section{Proposed Algorithm: Tracking-SHS}\label{sec4}

\noindent In the real world, it is improbable that the network evolves drastically within a short time. The similarity in the structure of the network before and after updates leads to a similar SHS set. We propose an efficient \textbf{Tracking-SHS algorithm}, that maintains Top-$k$ SHSs in the dynamic network. This algorithm considers the situations where it is crucial to handle the decremental updates in the network, such as outdated web links in web graphs or obsolete user profiles in a social network. As a result, the discovered SHSs change due to the dynamic nature of the network. Therefore, it is vital to design an algorithm that quickly discovers SHSs as the network evolves. Our proposed Tracking-SHS algorithm addresses this issue.

Instead of constructing the SHS set from the ground; we start from old Top-$k$ set and repeatedly update it. The goal of our algorithm is to maintain and update the SHS set faster than recomputing it from scratch. This section first discusses the mechanism
to find the set of affected nodes and various cases due to updates in the network. We also present a method to compute the PC score of the nodes efficiently and the procedure to update Top-$k$ SHSs for decremental edge updates in the network. Later, we extend our proposed single-edge update algorithm to a batch of updates.

\subsection{Finding Affected Nodes}\label{sec4.1}
\noindent Whenever there is an edge update in the network, by identifying the set of affected nodes, we need to recompute the pairwise connectivity scores of these nodes only.

 \begin{figure}[t!]
  \centering  \includegraphics[width=0.4\paperwidth]{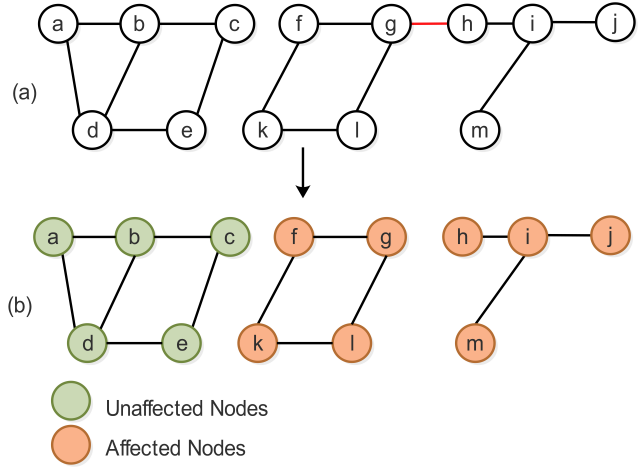}
    \caption{Illustration of affected and unaffected nodes due to updates in the network (a) Original network (b) Updated network.}
    \label{fig:aff}
  \end{figure}

\begin{definition}
\noindent \textbf{Affected Nodes.}  \normalfont Given an original graph $G(V,E)$, and an updated graph $G' = G(V, E\backslash (a,b))$, affected nodes are the set of nodes whose pairwise connectivities change as a result of the deletion of edge $(a, b)$. More precisely, the set of affected nodes $A$ is defined as $A = \{y \in V$ such that $c(y) \ne c'(y)\}$.
\end{definition}

\noindent When an edge $(a, b)$ is deleted from the network, affected nodes $A$ are the set of nodes reachable from node $a$, in case edge $(a, b)$ is a non-bridge edge. On the other hand, if edge $(a, b)$ is a bridge edge, we have two set of affected nodes $A_a$ and $A_b$, representing the set of nodes reachable from node $a$ and $b$, respectively. Let $G(V,E)$ be the original network, as shown in Figure \ref{fig:aff}(a), and Figure \ref{fig:aff}(b) shows the updated network $G'=G (V,E \backslash (g,h))$. We first show the case where edge deletion changes the PC score of some nodes $v$, i.e., $c(v) \ne c'(v)$. When an edge $(g, h)$ is deleted from $G$, the PC score of nodes $\{f, g, k, l, h, i, j, m\}$ changes as shown in Figure \ref{fig:aff}(b), and these nodes are called affected nodes. Next is the case where edge deletion does not change the PC score of the other set of nodes $u$, i.e., $c(u) = c'(u)$. Figure \ref{fig:aff}(b) shows that on the deletion of edge $(g,h)$, the PC score of the nodes $\{a, b, c, d, e\}$ does not change, and therefore, these nodes are unaffected nodes. On deletion of edge $(g, h)$ in Figure \ref{fig:aff}, set of affected nodes are $A_g = \{f, g, k, l\}$ and $A_h = \{h, i, j, m\}$.

\begin{lemma}
\label{l_1}
\normalfont \textit{\textbf{Given a graph $G = (V,E)$ and an edge update $(a, b)$, any node $v \in V$ is an affected node if $u(v, a) = 1$ or $u(v, b) = 1$, resulting in $c(v)$ before deletion not equal to $c'(v)$ after deletion. Otherwise, the node is unaffected.}}
\end{lemma}

\noindent \begin{proof}
Node $v$ is affected if it is either reachable from node $a$ or $b$ via any path in the graph. Nodes that are not reachable either from node $a$ or $b$ are not affected as these nodes do not contribute to the PC score of node $a$ or $b$. 
\end{proof}

\subsection{Various Cases for Edge Deletion}
\noindent This section enumerates various cases that may arise due to the deletion of an edge from the network.\\

\noindent \textbf{Case 1: No change in connected component (Non-bridge edge)}

\begin{figure*}[!t]
  \centering    \includegraphics[width=0.55\paperwidth]{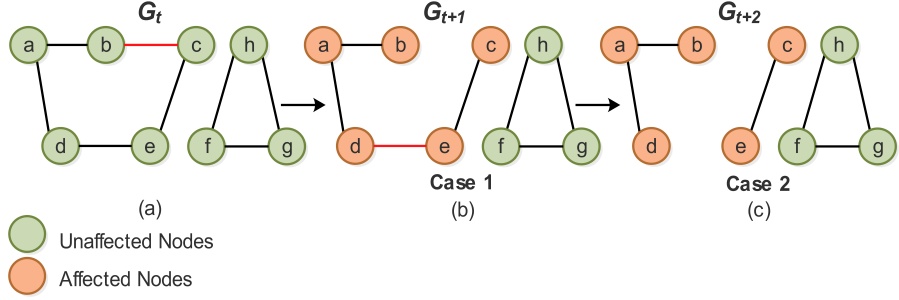}
    \caption{Illustration of cases after edge deletion.}
    \label{fig:cases}
  \end{figure*}

\noindent When the deleted edge $(a,b)$ is a non-bridge edge, there is no change in the connected components of the updated network, and node $a$ and $b$ still belong to the same connected component. The PC score of only few nodes in the component changes, i.e., $c(r) \neq c'(r), \, r\in C(a)$. Here, the maximum number of affected nodes are the nodes in the connected component containing node $a$, i.e., $ \mid A \mid $ = $ \mid C(a) \mid $. Let Figure \ref{fig:cases}(a) shows the original network $G_{t}$, when an edge $(b,c)$ is deleted at $G_{t+1}$, node $b$ and $c$ belong to the same connected component. However, the PC score of some nodes changes as highlighted in Figure \ref{fig:cases}(b).\\

\noindent \textbf{Case 2: Split connected component (Bridge edge)}

\noindent When edge $(a,b)$ is a bridge edge, its deletion splits the connected component into two new connected components. The PC score of the nodes in the component containing node $a$ and node $b$ changes, i.e., $c(r) \neq c'(r) \; \forall \,r \in C(a)$ or $C(b)$. Here, the number of affected nodes $ \mid A_a \mid $ and $ \mid A_b \mid $ are the number of nodes reachable from node $a$ and node $b$, respectively, and total affected nodes $ \mid A \mid  =  \mid A_a \mid  +  \mid A_b \mid $. Let Figure \ref{fig:cases}(b) be the original network at $G_{t+1}$. When edge $(d, e)$ is deleted at $G_{t+2}$, the connected component splits into two and the PC score of all the nodes in both the split components changes, as shown in Figure \ref{fig:cases}(c).

\subsection{Fast Computation of Pairwise Connectivity Score}
\noindent We have previously shown the method to calculate the PC score of the nodes. However, as the structure of the network changes, it is important to update the PC score of the affected nodes. PC score of node $i$ is computed as:
\begin{equation*}
   c(i)=P(G)-P(G\backslash\{i\}) 
\end{equation*}

Let $v \in R$ be a set of nodes not reachable from node $i$, i.e., $u(i, v) = 0\; \forall \,v \in R$. Therefore, nodes in $R$ do not contribute to the PC score of node $i$. Hence, it is not required to traverse the whole network to compute the PC score of node $i$, instead traversing the component to which node $i$ belongs is sufficient. This makes the PC score computing mechanism faster since we consider only the component to which the node belongs while ignoring the rest of the network. Updated PC score for node $i$ can be calculated as:
\begin{equation*}
c(i)=P(C(i))-P(C(i)\backslash\{i\})\\
\end{equation*}
\begin{equation}
\label{eq_5}
    c(i)=\tbinom{ \mid C(i) \mid }{2}-{\displaystyle \sum_{1\text{\ensuremath{\le}}j\text{\ensuremath{\le}}r}}\tbinom{ \mid C_j \mid }{2}\\
\end{equation}

Let node $i$ connects the component $C_1,..,C_r$, where $r$ denotes the number of distinct components containing neighbors of node $i$. The first term in Equation (\ref{eq_5}) gives the PC score of the component containing node $i$, and the second term gives the PC score of this component without node $i$. The difference between both the terms gives the PC score of node $i$. Here, $P(C(i))$ denotes the PC score of the component containing node $i$.

\subsection{Updating Top-$k$ Spanners}\label{sec4.4}

\noindent This section discusses the procedure for updating Top-$k$ SHSs. We use Algorithm \ref{alg 1} to obtain the initial SHS set and PC score of the nodes in the original network. In addition, we use max-heap priority queue $Q$, where the nodes are sorted by their PC score. The top node $w$ in the priority queue has the highest PC score $c(w)$, among all the nodes in the network. After every update, the PC scores in the priority queue $Q$ change according to Lemma \ref{l_2} and Lemma \ref{l_3}. Besides, we have maintained a min-heap priority queue for the SHS nodes in Top-$k$.

\begin{lemma}
\label{l_2}
\normalfont \textit{\textbf{Given a graph $G = (V, E)$ and an edge update $(a,b)$, if the deleted edge $(a,b)$ is a non-bridge edge, then $c'(v) \geq c(v)$, $\forall \,v \in A$.}}
\end{lemma}
\noindent \begin{proof}
Deletion of a non-bridge edge $(a,b)$ may bring some nodes of the graph into the bridging position, which leads to a higher PC score of these nodes compared to their previous PC score. For the rest of the nodes, their new PC score remains the same.
\end{proof}

\begin{lemma}
\label{l_3}
\normalfont \textit{\textbf{Given a graph $G = (V,E)$ and an edge update $(a, b)$, if the deleted edge $(a, b)$ is a bridge edge, then $c'(v) < c(v)$, $\forall \,v \in A$.}}
\end{lemma}
\noindent \begin{proof}
Deletion of a bridge edge $(a,b)$ splits the connected component into two components. In the updated graph, the nodes in the split connected components are now pairwise connected to a smaller number of nodes, due to which their updated PC score is less as compared to their previous PC score, i.e., $c'(v) < c(v)$, $\forall \, v \in A$.\end{proof}\\

\begin{algorithm}[t!]
\caption{Tracking-SHS algorithm: Updating Top-$k$ SHSs on deletion of an edge}
 \label{alg 2}
 \begin{algorithmic}[1]
 \renewcommand{\algorithmicrequire}{\textbf{Input:}}
 \renewcommand{\algorithmicensure}{\textbf{Output:}}
 \REQUIRE Graph $G(V, E)$,  Old SHS set Top-$k$,  Deleted edge $(a, b)$, Priority queue $Q$ with nodes sorted by PC score $c$ (The PC scores in $Q$ change according to Lemma \ref{l_2} and \ref{l_3})
 \ENSURE Updated SHS set Top-$k$
 \STATE Determine if edge $(a,b)$ is a bridge or non-bridge edge
 \STATE Identify set of affected nodes $A$
 \FORALL{$v\in A$}
   \STATE Compute $c'(v)$
   \STATE $Q(v)\leftarrow c'(v)$
 \ENDFOR
  
 \FORALL{$v\in$ Top-$k$}
   \STATE Compute $c'(v)$
 \ENDFOR
  
 \WHILE{$Q$ is not empty}
   \STATE $w\leftarrow Q.getMax()$
   \IF {$c(w)\leq $ Top-$k.getMin()$}
      \RETURN Top-$k$
   \ELSIF{$w\in$ Top-$k$}
     \STATE $G=G\backslash\{w\}$
     \STATE update $Q$
   \ELSE
     \STATE Top-$k.removeMin()$
     \STATE Top-$k.insert(c(w),w)$
     \STATE $G=G\backslash\{w\}$
     \STATE update $Q$
   \ENDIF
 \ENDWHILE
\end{algorithmic}
\end{algorithm}

Algorithm \ref{alg 2} presents the Tracking-SHS procedure for updating Top-$k$ SHSs. 
The algorithm works as follow. When an edge $(a,b)$ is deleted from the network, it is first determined if it is a bridge or non-bridge edge. We run DFS from $a$ and stop as we hit $b$. If we hit $b$, it indicates that edge $(a,b)$ is a non-bridge edge, otherwise bridge edge. We then identify the set of affected nodes using the procedure discussed in Section \ref{sec4.1}. We now compute the new PC score of the affected nodes using Equation (\ref{eq_5}) and update the PC score of these nodes into the priority queue $Q$. The PC score of the nodes in Top-$k$ may also change due to updates in the network, and therefore, we need to update the PC score of affected nodes in Top-$k$. Once we have updated the PC scores of all the nodes in the network, we update the Top-$k$ SHS set. 

Following Lemma \ref{l_2}, we know that the updated PC score of the affected nodes either increases or remains the same (in the case of non-bridge edge). Since the PC score of some nodes in the network may increase, we need to determine if any of these affected nodes have a higher PC score than existing SHS nodes in Top-$k$ and add such nodes in the SHS set. Besides, following Lemma \ref{l_3}, we know that the updated PC score of the affected nodes is always less compared to their previous PC score (in case the deleted edge is a bridge edge). Since the updated PC score of the affected nodes decreases, we need to replace the affected nodes present in Top-$k$ with the high PC score non-SHS nodes in the network.

Let $w$ denotes the node with the maximum PC score in $Q$. Now, we compare the PC score of $w$, i.e., $c(w)$ with the minimum PC score node in Top-$k$, and if $c(w) \leq $ Top-$k.getMin()$, we terminate the algorithm and return Top-$k$. In contrast, if $c(w) >$ Top-$k.getMin()$, and node $w$ is already present in Top-$k$, we remove $w$ from the network and update the PC score of the nodes in the component containing node $w$ in $Q$. On the other hand, if node $w$ is not present in Top-$k$, we remove the minimum PC score node from Top-$k$ and add node $w$ to Top-$k$. Finally, $w$ is removed from $G$, and the PC score of the nodes in the priority queue is updated.

\begin{lemma}
\label{l_4}
\normalfont \textit{\textbf{Given a graph $G =(V, E)$ and an edge update $(a, b)$, Algorithm \ref{alg 2} replaces a maximum of $(k-k')$ nodes from the Top-$k$ SHS set on the deletion of a bridge edge.}}
\end{lemma}

\noindent \begin{proof} Following Lemma \ref{l_3}, when the deleted edge $(a,b)$ is a bridge edge, the PC score $c(i) \, \forall \, i\in A$ decreases, whereas the PC score $c(i)$ $\forall$ $i \notin A$ remains the same. There may be some nodes that are affected and are present in Top-$k$ SHS set. Such nodes need to be replaced with high PC score non-SHS nodes from the network. Consider Top-$k_{rem}$, of size $ \mid $Top-$k_{rem} \mid  = k'$ as the set of nodes in Top-$k$ which are not affected, i.e., Top-$k_{rem}$ = Top-$k\backslash A$. Therefore, we have $(k-k')$ affected nodes in Top-$k$, and at most $(k-k')$ nodes in Top-$k$ can be replaced by high PC score non-SHS nodes in the network.\end{proof}

\subsection{Updating Top-$k$ SHSs for Batch Updates}
\noindent In the case of batch updates, a number of edges may be deleted from the network. 
One solution to update Top-$k$ SHSs for a batch update is to apply our proposed single edge update algorithm, i.e., Tracking-SHS after every update. However, this approach may not be efficient for a large number of updates. Therefore, we propose an efficient method that works as follows.

Let us consider that a set of batch updates consists of $l$ individual updates where $l\in(1,m)$. When a batch of $l$ edges is deleted from the network, each edge can either be a bridge edge or a non-bridge edge. We first determine the set of connected components in the network after the deletion of a batch of edges. Then, identify the set of affected nodes due to the updates in the network. For $l$ single updates, each update $i$ has its own affected node set $A_i,\, i = 1, 2, 3,...l$. For batch update, the total affected node set $A$ is the union of $A_i$, for all $i \in (1, l)$.
\begin{equation*}
    A=\bigcup_{i=1}^{l}A_{i}
\end{equation*}
Once we have an affected node set due to batch updates, the PC scores of the affected nodes are recomputed using the efficient PC score computing function (Equation (\ref{eq_5})). Finally, update Top-$k$ SHSs set using the procedure discussed in Section \ref{sec4.4}. In the case of a large number of updates, processing batch updates is more efficient than sequentially processing each update. For instance, if a node is affected several times during serial edge update, we need to recompute its PC score every time it is affected. However, in the case of a batch update, we need to recompute its PC score only once, making the batch update procedure more efficient.

\section{Proposed Model: GNN-SHS}\label{sec5}
\noindent Inspired by the recent advancement of graph neural network techniques on various graph mining problems, we propose \textbf{\textit{GNN-SHS}}, a graph neural network-based framework to discover Top-$k$ SHS nodes in the dynamic network. This model considers the situations where it is important to handle incremental as well as decremental updates in the network, such as Facebook, where links appear and disappear whenever a user friend/unfriend others. Due to dynamic nature of network, discovered SHSs change; therefore, it is crucial to design a model that can efficiently discover SHSs as the network evolves. Figure \ref{fig:gnn_arch} represents the architecture of the proposed GNN-SHS model. We divided the SHSs identification process into two parts, i.e., model training and model application. The details of the GNN-SHS model are discussed below.

\subsection{Model Training} \label{model_train}
\noindent This section discusses the architecture of the proposed model and the training procedure.

\subsubsection{Architecture of GNN-SHS} 
\noindent In order to discover SHS nodes in dynamic network, we first transform the SHSs identification problem into a learning problem. We then propose a GNN-SHS model that uses the network structure as well as node features to identify SHS nodes. Our model utilizes three-node features, i.e., effective size \cite{burt1992structural}, efficiency \cite{burt1992structural} and degree, to characterize each node. These features are extracted from the one-hop ego network of the node. 

\begin{figure*}[t!]
 \centering
\includegraphics[width=0.55\paperwidth]{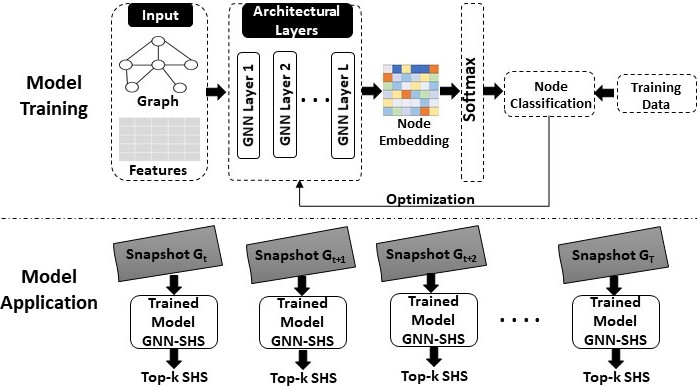}
 \caption{Architecture of proposed model GNN-SHS.}
 \label{fig:gnn_arch}
\end{figure*}

Given a graph and node features as input, our proposed model GNN-SHS first computes the low-dimensional node embedding vector and then uses the embedding of the nodes to determine the label of nodes (as shown in Figure \ref{fig:gnn_arch}). The label of a node can either be SHS or normal. The procedure for generating embeddings of the nodes is presented in Algorithm \ref{gnn-algo}. The model training is further divided into two phases: 1) Neighborhood Aggregation, 2) High Order Propagation. The two phases of the GNN-SHS model are discussed below:\\

\begin{algorithm}[ht!]
\caption{Generating node embeddings for GNN-SHS}
 \label{gnn-algo}
 \begin{algorithmic}[1]
 \renewcommand{\algorithmicrequire}{\textbf{Input:}}
 \renewcommand{\algorithmicensure}{\textbf{Output:}}
 \REQUIRE Graph $G(V,E)$, Input features $\vec{x}(i),\,\, \forall i \in V$, Depth $L$, Weight matrices $W^{l},\,\, \forall l \in \{1,..,L\}$, Non-linearity $\sigma$
\ENSURE Node embeddings $z{(i)}, \,\, \forall i \in V$ 
 \STATE $h^{0}(i) \leftarrow \vec{x}(i),\,\,\forall i \in V$ 
 \FOR{$l = 1$ to $L$}
 \FOR{$i \in V$}
  \STATE Compute $h^{l}{(N(i))}$ using Equation (\ref{eq:agg})
  \STATE Compute $h^{l}{(i)}$ using Equation (\ref{eq:comb})
 \ENDFOR
 \ENDFOR
 \STATE $z{(i)} = h^{L}{(i)}$
\end{algorithmic}
\end{algorithm}

\noindent \textbf{Neighborhood Aggregation.} 
The neighborhood aggregation phase aggregates the features from the neighbors of a node to generate node embeddings. The node embeddings are the low dimensional representation of a node. Due to distinguishing characteristics exhibited by the SHSs, we considered all the one-hop neighbors of the node to create embedding. We generate the embeddings of node $i$ by aggregating the embeddings from its neighboring nodes, and we use the number of neighbors of node $i$ as weight factor:
\begin{equation}
\label{eq:agg}
h^{l}{(N(i))} = \sum_{j\in N(i)}{\frac{h^{l-1}{(j)}}{ \mid N(i) \mid }}
\end{equation}
where $h^{l}{(N(i))}$ represents the embedding vectors captured from the neighbors of node $i$. Embedding vector of each node is updated after aggregating embeddings from its neighbors. Node embeddings at layer $0$, i.e., $h^{0}(i)$ are initialized with the feature vectors $\vec{x}(i)$ of the nodes, i.e., Effective size, Efficiency and Degree. Each node retains its own feature information by concatenating its embedding vector from the previous layer with the aggregated embedding of its neighbors from the current layer as:
\begin{equation}
\label{eq:comb}
h^{l}{(i)} = \sigma{\Big(W^{l}\big(h^{l-1}{(i)} \mathbin\Vert h^{l}{(N(i))}\big) \Big)}
\end{equation}
where $W^{l}$ are the training parameter, $\mathbin\Vert$ is the concatenation operator, and $\sigma$ is the non-linearity, e.g., ReLU. \\

\noindent \textbf{High Order Propagation.} Our model employs multiple neighborhood aggregation layers in order to capture features from $l$-hop neighbors of a node. The output from the previous layer acts as input for the current layer. Stacking multiple layers will recursively form the representation $h^{l}(i)$ for node $i$ at the end of layer $l^{th}$ as: 
\begin{equation*}
z{(i)} = h^{l}{(i)}, \,\,\,\, \forall i \in V
\end{equation*}
where $z(i)$ denotes the final node embedding at the end of $l^{th}$ layer. For the purpose of classifying the nodes as SHS or normal node, we pass the final embeddings of each node $z{(i)}$ through the Softmax layer. This layer takes node embeddings as input and generates the probability of two classes: SHS and normal. We then train the model to distinguish between SHS and normal nodes.

\subsubsection{Training Procedure} 
\noindent To discover SHS nodes in the network, we employ \textit{binary cross-entropy loss} to train the model with the actual label known for a set of nodes. The loss function $\mathcal{L}$ is computed as:

\begin{equation*}
\label{eq_loss}
\mathcal{L} = {-}\frac{1}{ r}{\sum_{i=1}^{r}\bigg(y(i)\log{\hat{y}(i)} + (1-y(i)) \log{(1-\hat{y}(i))\bigg)}}
\end{equation*}
where $y$ is the true label of a node, $\hat{y}$ is the label predicted by the GNN-SHS model, and $r$ is the number of nodes in the training data for which the labels are known.

\subsection{Model Application}
\noindent We first train the GNN-SHS model using the labelled data as discussed in Section \ref{model_train}. Once the model is trained, we then utilize the trained GNN-SHS model to quickly discover SHS nodes in each snapshot of graph (snapshot obtained from the dynamic network). The discovered SHSs are nodes removal of which minimizes the total pairwise connectivity in the remaining subgraph.

\begin{theorem} \label{thm2}
\textit{\textbf{The depth of the proposed graph neural network-based GNN-SHS model with \textit{width} $=O(1)$ should be at least $\Omega({n}^2/\log^2 n)$ to solve the SHSs problem.} 
}
\end{theorem}

\noindent \begin{proof}
In Theorem \ref{thm1}, we proved that if we can discover SHS nodes in the graph, then we can solve the VC problem. Corollary 4.4 of Loukas \cite{loukas2019graph} showed that for solving the minimum VC problem, a message-passing GNN with a \textit{ width} $=O(1)$  should have a depth of at least $\Omega({n}^2/\log^2 n)$. Therefore, the lower bound on the depth of the VC problem also applies to our SHSs problem. Here, the \textit{depth} describes the number of layers in the GNN-SHS model, and \textit{width} indicates the number of hidden units.
\end{proof}

\section{Theoretical Analysis of Tracking-SHS algorithm}\label{sec6}
\noindent This section presents the theoretical performance analysis of the proposed Tracking-SHS algorithm. We first analyze the run time of the single edge update algorithm for various cases discussed in Section \ref{sec4}. We then give an overall run time of the proposed algorithm. Later, we analyze the performance of the batch update algorithm.
When there are more connected components in the network, the time to update Top-$k$ SHSs will be less as there will be fewer affected nodes. In addition, the time to recompute the PC score of a node will also be less due to the small size of connected components. We consider the worst-case by assuming only one connected component in the initial network. We take the value of $k$ as 1. It takes $(n+m)$ time to determine if edge $(a,b)$ is a bridge or non-bridge. We run DFS to identify the affected node set, and the time required for the same is $(n+m)$. The PC score of a node is calculated using Equation (\ref{eq_5}). Notably, the time to update Top-$k$ SHSs is the time to recompute the PC scores of the affected nodes, i.e., $( \mid A \mid \times\, m)$, and a detailed analysis of $A$ and $m$ is shown below.

\begin{lemma}
\label{l_5}
\normalfont \textit{\textbf{Given a graph $G = (V,E)$, old SHS set Top-$k$ and an edge update $(a, b)$, Algorithm \ref{alg 2} takes $mn$ time to update the Top-$k$ SHS set (In case of non-bridge edge).}}
\end{lemma}
\noindent \begin{proof}
In this case, the number of affected nodes $ \mid A \mid  =  \mid C(a) \mid $ and since there is only 1 connected component in the graph, the number of edges $m(a)$ in the connected component containing node $a$ after an edge update $(a,b)$ is $(m-1)$ and the number of affected nodes $ \mid A \mid $ can be at most $n$. Therefore, the time $T_{ab}$ required to update the spanner set can be at most $m(a)\times  \mid A \mid $, i.e., $mn$. Here, $T_{ab}$ denotes the time required by the proposed algorithm Tracking-SHS to update Top-$k$ SHSs after an edge update $(a,b)$. \end{proof}

\begin{theorem}
\label{t_1}
\normalfont \textit{\textbf{Given a graph $G = (V,E)$, old SHS set Top-$k$ and an edge update $(a,b)$, Algorithm \ref{alg 2} takes $\frac{5mn}{8}$ time to update Top-$k$ SHS set in the expected case (In case of bridge edge).}}
\end{theorem}

\noindent \begin{proof}
Deletion of a bridge edge $(a,b)$ splits the connected component into two components, one containing node $a$ and the other containing node $b$. The probability of deletion of any edge from the graph is uniform, i.e., $\frac{1}{m}$. Then, the size of each connected component after deletion of edge $(a,b)$ is:
\begin{align*}
\begin{split}
    m(a) {}& = i\\
\end{split}\\
\begin{split}
    m(b) {}& = (m-1-i)
\end{split}
\end{align*}
where $\,\, 0\le\,i\,\text{\ensuremath{\le}}\,\frac{m-1}{2}$\\

Here, we assume that deletion of a bridge edge $(a,b)$ results in two components and the number of edges in each split component follow the uniform distribution. Using this assumption, the expected size $\bar{m}(a)$, $\bar{m}(b)$ of the connected components is computed as:
\begin{align*}
\begin{split}
    \bar{m}(a) {}& = \sum_{i=0}^{\frac{m-1}{2}}i \times \frac{2}{m+1}\\
                & = \frac{(m-1)}{4}
\end{split}
\end{align*}
\begin{align*}
\begin{split}
    \bar{m}(b) {}& = \sum_{i=0}^{\frac{m-1}{2}}(m-1-i) \times \frac{2}{m+1}\\
                & = \frac{3(m-1)}{4}
\end{split}
\end{align*}

After an edge update $(a,b)$, the affected nodes are the set of nodes reachable from node $a$ and $b$. We use the property that the maximum number of nodes that may span $i$ edges in one connected component are $(i+1)$ to compute the number of affected nodes. The expected number of affected nodes $ \mid  \bar{A}_a \mid $, $ \mid \bar{A}_b \mid $ in both the components is calculated as:
\begin{align*}
\begin{split}
     \mid \bar{A}_a \mid  {}& = \sum_{i=0}^{\frac{m-1}{2}}(i+1)\times \frac{2}{m+1}\\
                & = \frac{(m-1)}{4}+1
\end{split}\\
\begin{split}
     \mid \bar{A}_b \mid  {}& = \sum_{i=0}^{\frac{m-1}{2}}(n-i-1)\times \frac{2}{m+1}\\
                & = n-(\frac{m-1}{4})-1
\end{split}
\end{align*}

To compute the number of affected nodes, we take the upper bound on the number of edges. Using the same property, we know that $m$ edges can be spanned by a maximum of $(m+1)$ nodes, but we have a maximum of $n$ nodes in the graph.\\

\noindent We put $m=n-1$ to compute the expected number of affected nodes.
\begin{align*}
\begin{split}
     \mid \bar{A}_a \mid  {}& = \frac{n-2}{4}+1\\
\end{split}\\
\begin{split}
     \mid \bar{A}_b \mid  {}& = n-(\frac{n-2}{4})-1\\
\end{split}
\end{align*}

\noindent Therefore, the expected time $\bar{T}_{ab}$ required to update the SHS set for this case is:
\begin{align*}
\begin{split}
    \bar{T}_{ab} {}& =  \mid \bar{A}_a \mid \times \,\bar{m}(a) +  \mid \bar{A}_b \mid \times \,\bar{m}(b)\\
\end{split}\\
\begin{split}
    \bar{T}_{ab} {}& = \frac{5mn}{8}\\
\end{split}
\end{align*}
Hence proved. \end{proof}

\begin{theorem}
\label{t_2}
\normalfont \textbf{\textit{For the general case, the overall time ${T}_{ab}$ required by the proposed algorithm to update Top-$k$ SHSs after an edge update $(a,b)$ is $P_{ab}(\frac{5mn}{8})+(1-P_{ab})(mn)$.}}
\end{theorem}

\noindent \begin{proof} Let $P_{ab}$ be the probability that edge $(a,b)$ is a bridge edge, removal of which splits the connected component into two components (Case 2), then the probability of an edge being a non-bridge is $(1-P_{ab})$ (Case 1). Using the results of Lemma \ref{l_5} and Theorem \ref{t_1}, the overall time for updating Top-$k$ SHSs after an edge update $(a,b)$ is given by ${T}_{ab}$ = $P_{ab}(\frac{5mn}{8})+(1-P_{ab})(mn)$.\end{proof}\\

\noindent The following special case stems from Theorem \ref{t_2}. \\

\noindent \textbf{\textit{Special case.}} For the bridge-edge dominating graph where all the edges in the graph are bridges, the probability of an edge being a bridge is 1, i.e., $P_{ab} = 1$ and non-bridge edge is $0$. Theoretically, the speedup of the proposed algorithm is $P_{ab}(\frac{5mn}{8})+(1-P_{ab})(mn)$ and substituting the values of $P_{ab}$ and $(1-P_{ab})$ gives us the overall update time, i.e., $T_{ab} = \frac{5mn}{8}$. In contrast, the time required for static recomputation is $T'_{ab} = (kn(m+n))$, i.e., $mn$ where $k$ is constant. Therefore, the proposed algorithm achieves an overall speedup of 1.6 times (speedup = $\frac{T'_{ab}}{T_{ab}}$) over recomputation. An example of such a graph is the Preferential Attachment graph. We use experimental analysis to validate our theoretical results, and the experimental results support our arguments.

\begin{theorem}
\label{t_3}
\normalfont \textit{\textbf{Given a graph $G = (V,E)$, old SHS set Top-$k$, and a batch of $l$ edge updates, the batch update algorithm, i.e., Tracking-SHS algorithm for batch updates takes $(N\_CC \times S\_CC \times A\_CC \times P_{l})$ time to update the Top-$k$ SHS set. Here, $N\_CC$ refers to the number of connected components in the resulting graph, $A\_CC$ is the number of affected nodes in the connected component, $S\_CC$ is the size of the connected component, and $P_l$ is the probability that deletion of $l$ edges results in $N\_CC$ in the updated graph.}}
\end{theorem}
\noindent \begin{proof}
The proof is straightforward and thus omitted. In Section \ref{sec7.3}, we will show that our batch update algorithm produces good results over static recomputation. \end{proof}

\section{Experimental Analysis}\label{sec7}

\noindent This section analyzes the performance of the proposed Tracking-SHS algorithm and GNN-SHS model. We first discuss the datasets used to evaluate the performance. We then evaluate the performance of the Tracking-SHS algorithm for single update followed by the batch of updates. Later, we discuss the performance of GNN-SHS model. We implemented our algorithms in Python 3.7. The experiments are performed on a Windows 10 PC with CPU 3.20 GHz and 16 GB RAM.

\subsection{Datasets} 

\noindent We measure the update time of the proposed Tracking-SHS algorithm, static recomputation and GNN-SHS model by conducting extensive experiments on various real-world and synthetic datasets. Due to computational challenges while computing the pairwise connectivity score of the nodes, it is not feasible to scale beyond the considered graph size using normal system configurations. The details of the datasets are discussed below.

\subsubsection{Real-World Datasets}

\noindent {To ensure consistency with prior studies in SHS detection, we selected widely recognized benchmark datasets, including Karate, Dolphins, Football, and HC-BIOGRID, which are standard in SHS research. These datasets allow for meaningful comparisons across studies and provide foundational insights into the efficiency of our methods in detecting structural hole spanners. We analyzed the performance of the proposed algorithms on these four real-world networks, each varying in size and structure. Specifically, the Karate dataset \cite{zachary1977information} represents a friendship network among members of a karate club; the Dolphin dataset \cite{lusseau2003bottlenose} represents frequent social associations between 62 dolphins in a community; the American College Football dataset \cite{girvan2002community} captures a network of games between Division IA colleges; and HC-BIOGRID\footnote{\url{https://www.pilucrescenzi.it/wp/networks/biological/}} represents a biological interaction network. The characteristics of these real-world datasets are summarized in Table \ref{dataset1}}.

\begin{table}[t!] 
\caption{Summary of real-world datasets.}
\label{dataset1}
\renewcommand{\arraystretch}{1.4}
\centering 
\begin{tabular}{llll} \hlineB{1.5} 
\textbf{Dataset} & \textbf{Nodes} & \textbf{Edges} & \textbf{Avg degree} \\ \hlineB{1.5}
Karate & 34 & 78 & 4.59 \\ 
Dolphins & 62 & 159 & 5.13\\ 
Football & 115 & 613 & 10\\ 
HC-BIOGRID & 4039 & 14342 & 7\\ \hlineB{1.5}
\end{tabular} 
\end{table}
\subsubsection{Synthetic Datasets}

\noindent  We analyze the performance of proposed algorithms by conducting experiments on synthetic datasets. We generate synthetic networks using graph-generating algorithms and vary the network size to determine its effect on algorithms performance. We conduct experiments on synthetic networks with diverse topologies: Preferential Attachment (PA) networks and Erdos-Renyi (ER) \cite{erdHos1959random} networks. We generate PA$(n)$ network with 500, 1000 and 1500 nodes, where $n$ denotes total nodes in the network. For ER$(n,p)$, we generate networks with 250 and 500 nodes, where $p$ is the probability of adding an edge to the network. In PA$(n)$ network, a highly connected node is more likely to get new neighbors. In ER$(n,p)$ network, parameter $p$ acts as a weighting function, and there are higher chances that the graph contains more edges as $p$ increases from 0 to 1. The properties of synthetic datasets are presented in Table \ref{dataset2}.

\begin{table}[ht!] 
\caption{Summary of synthetic datasets.}
\label{dataset2}
\renewcommand{\arraystretch}{1.4}
\centering 
\begin{tabular}{llll} \hlineB{1.5} 
\textbf{Dataset} & \textbf{Nodes} & \textbf{Edges} & \textbf{Avg degree} \\ \hlineB{1.5}
PA (500) & 500 & 499 & 2 \\ 
PA (1000) & 1000 & 999 & 2\\ 
PA (1500) & 1500 & 1499 & 2\\  
ER (250, 0.01) & 250 & 304 & 2\\ 
ER (250, 0.5) & 250 & 15583 & 124\\ 
ER (500, 0.04) & 500 & 512 & 2\\ 
ER (500, 0.5) & 500 & 62346 & 249\\ \hlineB{1.5} 
\end{tabular} 
\end{table}

\subsection{Performance of Tracking-SHS algorithm on Single Update}
\subsubsection{Performance on Real-World Dataset}

\noindent To evaluate the  performance of the Tracking-SHS algorithm for single edge update, we compare it against static recomputation. Table \ref{result_dataset1} shows the speedup achieved by the Tracking-SHS algorithm over recomputation. Speedup is the ratio of the speed of the static algorithm to that of Tracking-SHS algorithm, which is proportional to the ratio of computation time used by the static algorithm to that by the proposed Tracking-SHS algorithm. In order to determine how the two algorithms (static algorithm and proposed dynamic Tracking-SHS algorithm) perform for the dynamic network, we start with a  full network and randomly remove 50 edges, one at a time.
We then compute the geometric mean of the speedup for the proposed Tracking-SHS algorithm in terms of its execution time against the static algorithm. For instance, if the Tracking-SHS algorithm takes 10 seconds to execute for an edge update, whereas the static algorithm takes 50 seconds to execute for the same update, we say that the Tracking-SHS algorithm is 5 times faster than static recomputation. The column ``Gmean" has the geometric mean of the achieved speedup (over 50 edge deletions), ``Min" contains minimum achieved speedup and, ``Max" contains maximum speedup. We run our Tracking-SHS algorithm for 3 different values of $k$, i.e., $k$ = 1, 5 and 10. For real-world networks, the gmean speedup is always at least 2.35 times for $k$ = 1, 3.92 for $k$ = 5 and 5.02 for $k$ = 10.  The experimental results demonstrate that speedup increases with the value of $k$. The average speedup reaches 21.79 times for $k$ =10 from a speedup of 3.76 for $k$ = 1 (HC-BIOGRID dataset), which shows a significant improvement for a larger value of $k$. The average speedup over all tested datasets is 3.24 times for $k$ = 1, 6.56 for $k$ = 5, and 10.91 for $k$ = 10. The minimum speedup achieved by the proposed Tracking-SHS algorithm is 1.73 times for the Karate dataset, and the maximum speedup achieved is 22.65 times for HC-BIOGRID dataset. In addition, it has been observed that the speedup also increases with the size of the network. For a small size network (Karate dataset), the speedup is 5.02 times for $k$ = 10. In contrast, for the same value of $k$, the speedup increases significantly to 21.79 times for the large size network (HC-BIOGRID dataset).

\begin{table*}[t!] 
\begin{center}
\caption{Speedup of Tracking-SHS algorithm on static recomputation over 50 edge deletions on real-world datasets.}
\label{result_dataset1}
\renewcommand{\arraystretch}{1.4}
\centering 
\begin{tabular}{l|ccc|ccc|ccc} \hlineB{2.5} 
\textbf{Dataset} & \multicolumn{3}{c}{\textbf{\textit{k} = 1}} & \multicolumn{3}{|c}{\textbf{\textit{k} = 5}}& \multicolumn{3}{|c}{\textbf{\textit{k} = 10}}\\ \hlineB{1.5}
& \textbf{Gmean} & \textbf{Min} & \textbf{Max} & \textbf{Gmean} & \textbf{Min} & \textbf{Max} & \textbf{Gmean} & \textbf{Min} & \textbf{Max}\\ \hlineB{1.5}
Karate & 2.35 & 1.73 & 3.1 & 3.92 & 2.98 & 4.18 & 5.02 & 4.98 & 5.17 \\  
Dolphins & 3.34 &  2.11 & 4.18 & 4.16 & 3.06 & 5.33 & 7.52 & 5.21 & 9.22 \\  
Football & 3.72 & 3.42 & 4.21 & 10.17 & 9.6 & 11.47 & 17.26 & 15.45 & 19.84 \\  
HC-BIOGRID & 3.76 & 1.85 & 4.11 & 11.16 & 10.21 & 11.89 & 21.79 & 20.23 & 22.65 \\ \hlineB{1.5} 
\textbf{Mean (Geometric)} & \textbf{3.24} & \textbf{2.19} & \textbf{3.87} & \textbf{6.56} & \textbf{5.47} & \textbf{7.42} & \textbf{10.91} & \textbf{9.49} & \textbf{12.1}\\ \hlineB{1.5}
\end{tabular} 
\end{center}
\end{table*}

\begin{table*}[t!]
\caption{Speedup of Tracking-SHS algorithm on static recomputation over 50 edge deletions on synthetic datasets.}
\label{result_dataset2}
\renewcommand{\arraystretch}{1.4}
\centering 
\begin{tabular}{l|lll|lll|lll} \hlineB{1.5}
\textbf{Dataset} & \multicolumn{3}{c}{\textbf{\textit{k} = 1}} & \multicolumn{3}{|c}{\textbf{\textit{k} = 5}}& \multicolumn{3}{|c}{\textbf{\textit{k} = 10}}\\ \hlineB{1.5}
& \textbf{Gmean} & \textbf{Min} & \textbf{Max} & \textbf{Gmean} & \textbf{Min} & \textbf{Max} & \textbf{Gmean} & \textbf{Min} & \textbf{Max}\\ \hlineB{1.5}
PA (500) & 3.24 & 2.75 & 3.39 & 4.76 & 4.44 & 5.65 & 6.35 & 5.79 & 7.24 \\ 
PA (1000) & 3.32 & 3.08 & 3.83 & 5.3 & 4.46 & 6.12 & 8.31 & 7.75 & 9.31 \\ 
PA (1500) & 3.54 & 3.22 & 4.02 & 5.46 & 4.57 & 6.13 & 8.95 & 7.54 & 10.6\\ 
ER (250, 0.01) & 4.22 & 3.98 & 4.65 & 6.15 & 5.82 & 6.62 & 10.65 & 10.1 & 11.25 \\
ER (250, 0.5) & 4.08 & 3.9 & 4.17 & 11.66 & 10.42 & 13.15 & 20.4 & 18.34 & 22.21 \\ 
ER (500, 0.04) & 4.41 & 4.02 & 4.88 & 7.83 & 5.08 & 9.51 & 9.81 & 8.06 & 11.78 \\
ER (500, 0.5) & 3.95 & 3.78 & 4.35 & 11.39 & 10.33 & 12.52 & 21.44 & 20.35 & 29.08 \\  \hlineB{1.5}
\textbf{Mean (Geometric)} & \textbf{3.8} & \textbf{3.5} & \textbf{4.16} & \textbf{7.07} & \textbf{6.02} & \textbf{8.05} & \textbf{11.16} & \textbf{10.04} & \textbf{12.95}\\ \hlineB{1.5}
\end{tabular} 
\end{table*}

\subsubsection{Performance on Synthetic Dataset} 
\noindent We perform the similar experiments on diverse synthetic datasets of varying scales. We start with a full network and randomly remove 50 edges, one at a time. Table \ref{result_dataset2} shows the speedup of the proposed Tracking-SHS algorithm over static recomputation. For synthetic datasets, the gmean of the achieved speedup is always at least 3.24 times for $k$ = 1, 4.76 for $k$ = 5 and 6.35 for $k$ = 10 (over 50 edge deletions). Similar to the real-world dataset, the speedup increases with the increase in the value of $k$ for the synthetic dataset. For instance, in ER (500, 0.5) network, the average speedup increases to 21.44 times for $k$ = 10 from a speedup of 3.95 for $k$ = 1. The average speedup over all tested datasets is 3.8 for $k$ = 1, 7.07 for $k$ = 5, and 11.16 for $k$ = 10. Besides, speedup for ER dataset is relatively higher than the PA dataset of the same scale. Take an example of the PA dataset of 500 nodes, the mean speedup is 6.35 times, whereas, for ER dataset of 500 nodes, the mean speedup is at least 9.81 for $k$=10.

\subsection{Performance of Tracking-SHS algorithm on Batch Update}\label{sec7.3}
\begin{figure}[h!]
  \centering
    \includegraphics[width=0.39\paperwidth]{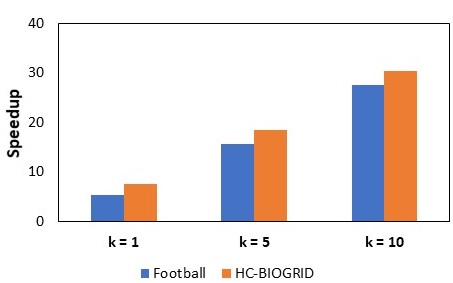}
    \caption{Speedup of Tracking-SHS algorithm on batch update.}
    \label{fig:batch}
  \end{figure}

\noindent We now demonstrate the performance of the batch update algorithm, i.e., Tracking-SHS algorithm for batch updates. Initially, we consider all the edges are present in the network and compute the Top-$k$ SHSs using Algorithm \ref{alg 1}. We then randomly remove a set of 50 edges from the network. These edges are considered as a batch of updates, and our goal is to update Top-$k$ SHS set corresponding to these updates. To evaluate the performance of our batch update algorithm, we performed experiments on the football and HC-BIOGRID dataset, and set the value of $k$ to 1,5 and 10. Figure \ref{fig:batch} shows the speedup achieved by the Tracking-SHS algorithm on batch updates.  We attained a speedup of 5.29 times for the football dataset and 7.51 times for the HC-BIOGRID dataset (for $k$ =1). Experimental results demonstrate that for batch update, the speedup increases with the increase in the value of $k$, e.g., for HC-BIOGRID, speedup is 7.51 times for $k$ =1, whereas it is 30.35 times for $k$ =10.

\subsection{Performance of GNN-SHS model}
\noindent In a dynamic network, as the graph changes with time, we get multiple snapshots of the graph. The trained model can be used to identify the SHSs in each snapshot of the dynamic graph. Even if it takes some time to train the model, we need to train it only once, and after that, whenever the graph changes, the trained GNN-SHS model can be used to discover the updated SHSs in a few seconds.

\subsubsection{Baseline}
\noindent In the literature, no solution addresses the problem of discovering SHSs in dynamic networks. Therefore, we compare the performance of our model GNN-SHS with the proposed Tracking-SHS algorithm (as discussed in Section \ref{sec4}).

\subsubsection{Evaluation Metrics}
\noindent We measure the \textit{efficiency} of our proposed GNN-SHS model in terms of speedup achieved by GNN-SHS over the Tracking-SHS algorithm. The speedup is computed as follows:
\begin{equation*}
\label{agg}
\text{Speedup} = \frac{\text{Run time of Tracking-SHS algorithm}}{\text{Run time of proposed GNN-SHS model}}
\end{equation*}

In addition, we measure the \textit{effectiveness} of our model in terms of classification accuracy achieved by the model. 

\subsubsection{Ground Truth Computation} 
\noindent To compute the ground truth labels, we first calculate the pairwise connectivity score $c$ using Equation (\ref{pc_score}) for each node and then label Top-$k$ nodes with the highest score as SHS nodes and rest as normal nodes. For experimental analysis, we set the value of $k$ to 50. 

\subsubsection{Training Settings} 
\noindent We implemented the code of GNN-SHS in PyTorch and fixed the number of layers to 2 and the embedding dimension to 32. The model parameters are trained using Adam optimizer, learning rate of 0.01 and weight decay of $5e-4$. We train GNN-SHS model for 200 epochs. We used 60\% of the nodes for training, 20\% for validation, and 20\% for testing. In addition, we used an inductive setting where test nodes are unseen to the model during the training phase\footnote{
{The GNN-SHS model used in this paper was intentionally designed as a lightweight architecture to balance accuracy and computational efficiency. This streamlined design allows for effective SHS detection in dynamic networks, particularly in environments with limited computational resources. However, the model is scalable and can be extended with additional layers or parameters for more complex applications, providing adaptability for a range of real-world dynamic graph settings.}}. 

\begin{table}[t!] 
\caption{Classification accuracy of GNN-SHS on synthetic datasets.}
\label{ACC_GNN-SHS}
\renewcommand{\arraystretch}{1.2}
\centering 
\begin{tabular}{ll} \hlineB{1.5}
\textbf{Dataset} & \textbf{Accuracy}\\ \hlineB{1.5}
PA (500) & 94.5\%\\ 
PA (1000) & 95.33\% \\ 
PA (1500) & 96.5\%\\ 
ER (250, 0.01) & 86\%\\
ER (250, 0.5) & 90\%\\ 
ER (500, 0.04) & 87\%\\
ER (500, 0.5) & 92\%\\ \hlineB{1.5}
\end{tabular} 
\end{table}

\subsubsection{Performance of GNN-SHS on Synthetic Dataset}
\noindent Tracking-SHS algorithm works for a single edge deletion update. Therefore, we analyze GNN-SHS performance on a single edge deletion update only so that we can compare the speedup of GNN-SHS over the Tracking-SHS algorithm. To determine the speedup of the proposed GNN-SHS model over the proposed Tracking-SHS algorithm, we start from the whole network and arbitrarily delete 50 edges; we only delete one edge at a time. In this way, we obtain multiple snapshots of the graph. We set the value of $k$ (number of SHS) to 50 and make use of the trained GNN-SHS model to discover SHS nodes in each new snapshot graph. We calculate the geometric mean of the speedup achieved by GNN-SHS over the Tracking-SHS algorithm. \\

Table \ref{ACC_GNN-SHS} reports the high classification accuracy achieved by our GNN-SHS model on various synthetic graphs. With a minimum accuracy of 94.5\% on the Preferential Attachment graph PA(500) and 86\% for the Erdos-Renyi graph ER(250, 0.01), our model consistently demonstrates its effectiveness in SHS classification.

\begin{table}[t!] 
\caption{Speedup of GNN-SHS model on Tracking-SHS algorithm over 50 edge deletions on synthetic datasets.}
\label{SPEED_GNN-SHS}
\renewcommand{\arraystretch}{1.2}
\centering 
\begin{tabular}{llll} \hlineB{1.5}
\textbf{Dataset} & \textbf{Geometric Mean} & \textbf{Min} & \textbf{Max} \\ \hlineB{1.5}
PA (500) & 1236.4 & 1012.5 & 1532.7\\ 
PA (1000) & 1930.6 & 1574.2 & 2141.4\\ 
PA (1500) & 2639.7 & 2432.1 & 2996.9\\ 
ER (250, 0.01) & 37.5 & 31.8 & 40.2\\
ER (250, 0.5) & 287.3 & 263.6 & 301.2\\ 
ER (500, 0.04) & 368.2 & 354.5 & 379.3\\
ER (500, 0.5) & 2466.6 & 2015.3 & 2845.9\\ \hlineB{1.5}
\textbf{Mean (Geometric)} & 671.6 & 584.1 & 745.9\\ \hlineB{1.5}
\end{tabular} 
\end{table}

Table \ref{SPEED_GNN-SHS} reports the speedup achieved by the proposed GNN-SHS model over the Tracking-SHS algorithm. \textit{GNN-SHS model achieved high speedup over the Tracking-SHS algorithm while sacrificing a small amount of accuracy.} \textit{The proposed model GNN-SHS is at least 31.8 times faster for the ER(250, 0.01) network and up to 2996.9 times faster for PA(1500) over the Tracking-SHS algorithm, providing a considerable efficiency advantage.} The geometric mean speedup is always at least 37.5 times, and the average speedup over all tested datasets is 671.6 times. Results show that our graph neural network-based model GNN-SHS speeds up the SHS identification process in dynamic networks. In addition, it has been observed from the results that the speedup increases as network size increases, e.g., for PA graphs, the geometric mean speedup is 1236.4 times for a graph of 500 nodes, 1930.6 times for a graph with 1000 nodes and 2639.7 times for a graph with 1500 nodes.

In Theorem 2, we showed that the depth of GNN-SHS should be at least $\Omega({n}^2/\log^2 n)$ to solve the SHSs problem. Nevertheless, a deeper graph neural network suffers from an over-smoothing problem \cite{li2018deeper, yang2020toward}, making it challenging for GNN-SHS to differentiate between the embeddings of the nodes. In order to avoid the over-smoothing problem, we only used two layers in our GNN-SHS model.

\begin{table}[bh!] 
\caption{Run time (sec) and classification accuracy of GNN-SHS model on real-world datasets.}
\label{result-real}
\renewcommand{\arraystretch}{1.4}
\centering 
\begin{tabular}{p{2cm}ll} \hlineB{1.5} 
\textbf{Dataset} & \textbf{Run time (sec)}& \textbf{Accuracy} \\ \hlineB{1.5}
Dolphin & 0.002 & 76.92\% \\ 
Football  & 0.009 & 86.96\%\\ 
 \hlineB{1.5} 
\end{tabular} 
\end{table}

\subsubsection{Performance of GNN-SHS on Real-world Dataset}
\noindent We perform experiments on real-world datasets to determine the proposed model GNN-SHS performance for incremental and decremental batch updates. In the literature, no solution discovers SHSs for incremental and decremental batch updates; therefore, we can not compare our results with other solutions. We only report the results obtained from our experiments. We set the value of $k=5$ (number of SHSs). For each real-world dataset, we initiate with the whole network and then arbitrarily delete five edges from the network and add five edges to the network at once. In this manner, we obtain a snapshot of the graph. We then use our trained model GNN-SHS to discover SHSs in the new snapshot graph. Our empirical results in Table \ref{result-real} show that our model discovers updated SHSs in less than 1 second for both Dolphin and American College Football datasets. Besides, our model achieves high classification accuracy in discovering SHSs for batch updates.

{Notably, GNN-SHS demonstrates strong performance in dynamically changing networks; however, it is essential to acknowledge the potential effects of data drift over time. As the model is trained on specific network snapshots, significant structural changes in the network could affect the accuracy of its SHS predictions. This phenomenon, common in machine learning models, occurs when shifts in the underlying data distribution (network structure) impact model performance. As networks evolve, particularly in cases with substantial and continuous structural updates, the learned node embeddings may no longer accurately capture the dynamics necessary for effective SHS detection. To address this, periodic retraining of GNN-SHS with updated network snapshots may be beneficial, especially for long-term applications with frequent incremental and decremental changes. This approach can mitigate the effects of data drift, ensuring that GNN-SHS remains robust and accurate as the network evolves. Our experiments on synthetic and real-world datasets demonstrate that GNN-SHS can maintain high classification accuracy across multiple snapshots without frequent retraining, which makes it an efficient choice for scenarios with moderate changes. However, for use cases involving rapidly evolving network structures, periodic retraining represents a practical compromise that sustains the model’s adaptability and performance in dynamic settings. This selective retraining approach allows GNN-SHS to balance efficiency and adaptability, optimizing performance based on the specific demands of the network environment. For real-time applications, where computational efficiency is essential, the model can operate effectively with minimal retraining. In contrast, in high-volatility scenarios, periodic retraining enhances adaptability to maintain prediction accuracy across diverse dynamic conditions.}

\section{Discussion}

{In this study, we focus on foundational dynamic updates, specifically single-edge and batch-edge additions and deletions, to establish baseline insights into SHS detection efficiency in evolving networks. While real-world networks often exhibit more complex dynamics, such as community evolution or node attribute changes, our work provides a critical foundation for exploring these advanced scenarios. The proposed approaches, Tracking-SHS and GNN-SHS, contribute significantly to the field of SHS detection, employing both algorithmic and machine learning methods to identify structural hole spanners in dynamic networks. The Tracking-SHS algorithm achieves 100\% accuracy through deterministic computation, ensuring complete precision in SHS identification for decremental updates. This makes it a valuable tool for scenarios that require absolute accuracy. However, the GNN-SHS model balances high accuracy with substantial computational speedup. Although it does not achieve 100\% accuracy, it consistently delivers strong performance, offering significant advantages in time-sensitive applications. This trade-off between precision and efficiency is especially beneficial for evolving networks, where rapid SHS detection can yield actionable insights even without absolute precision. This study also employs widely recognized benchmark datasets and synthetic networks, providing a robust foundation for evaluating SHS detection methods. While these datasets establish baseline performance, future research could explore the scalability and adaptability of the proposed methods in larger, more intricate real-world networks. Moreover, the GNN-SHS model is trained once to maintain computational efficiency in environments with moderate structural changes. For more volatile networks requiring higher adaptability, selective retraining could enhance the model's flexibility. By emphasizing fundamental dynamic changes, our approach provides a clear understanding of SHS detection efficiency while setting the stage for investigating more complex network dynamics in future work. }

\section{Conclusion and Future Work}\label{sec8}
\noindent The structural hole spanner discovery problem for various applications, including community detection, viral marketing, etc. However, the problem has not been studied for dynamic networks. In this paper, we studied the SHS discovery problem for dynamic networks. We first proposed an efficient Tracking-SHS algorithm that maintains SHSs dynamically by discovering an affected set of nodes whose connectivity score updates as a result of changes in the network. We proposed a fast procedure for calculating the scores of the nodes. We also extended our proposed single edge update Tracking-SHS algorithm to a batch of edge updates. In addition, we proposed a graph neural network-based model, GNN-SHS, that discovers SHSs in dynamic networks by learning low-dimensional embedding vectors of nodes. Finally, we analyzed the performance of the Tracking-SHS algorithm theoretically. We showed that our proposed algorithm achieves a speedup of 1.6 times over recomputation for a particular type of graph, such as Preferential Attachment graphs. In addition, our experimental results demonstrated that the Tracking-SHS algorithm is at least 3.24 times faster than the recomputation with a static algorithm, and the proposed GNN-SHS model is at least 31.8 times faster than the comparative method, demonstrating a considerable advantage in run time.

{In future work, we aim to investigate the scalability of our methods on larger and more diverse dynamic networks by incorporating additional real-world datasets to validate robustness. Extending our algorithms to handle directed graphs, weighted edges, and advanced dynamics, such as community evolution, could enhance adaptability. Furthermore, periodic retraining for GNN-SHS can improve performance in rapidly evolving settings. Expanding comparisons with state-of-the-art methods and applying the approach in practical scenarios, such as social network analysis and cybersecurity, would provide valuable insight into its utility in various domains.}

\bibliographystyle{ACM-Reference-Format}
\bibliography{sn-bibliography}

\end{document}